\documentclass[prl,twocolumn,unsortedaddress,showpacs]{revtex4}
\usepackage{graphicx}
\usepackage{amssymb}
\usepackage{bm}
\begin{document}
\setcounter{page}{1}
\title[]{Impact of in-plane currents on magnetoresistance properties of an
exchange-biased spin-valve with insulating antiferromagnetic layer}
\author{D. N. H. \surname{Nam}}
\thanks{Corresponding author}
\email{daonhnam@yahoo.com}
\affiliation{Institute of Materials Science, VAST, 18 Hoang-Quoc-Viet, Hanoi,
Vietnam}
\author{N. V. \surname{Dai}}
\affiliation{Institute of Materials Science, VAST, 18 Hoang-Quoc-Viet, Hanoi,
Vietnam}
\affiliation{q-Psi and Department of Physics, Hanyang University, Seoul
133-791, Korea}
\author{N. C. \surname{Thuan}}
\author{L. V. \surname{Hong}}
\author{N. X. \surname{Phuc}}
\affiliation{Institute of Materials Science, VAST, 18 Hoang-Quoc-Viet, Hanoi,
Vietnam}
\author{S. A. \surname{Wolf}}
\affiliation{Department of Materials Science and Engineering, University of
Virginia, Charlottesville, Virginia 22903, USA}
\author{Y. P. \surname{Lee}}
\affiliation{q-Psi and Department of Physics, Hanyang University, Seoul
133-791, Korea}

\date[]{Received \today}

\begin{abstract}
The impact of in-plane alternating currents on the exchange bias,
resistance, and magnetoresistance of a
Co$_{85}$Fe$_{15}$/Ni$_{0.85}$Co$_{0.15}$O/Co$_{85}$Fe$_{15}$/Cu/Co$_{85}$Fe$_{15}$
spin-valve is studied. With increasing current, the resistance is
increased while the maximum magnetoresistance ratio decreases. Noticeably,
the reversal of the pinned layer is systematically suppressed in both field
sweeping directions. Since Ni$_{0.85}$Co$_{0.15}$O oxide is a good
insulator, it is expected that the ac current flows only in the
Co$_{85}$Fe$_{15}$/Cu/Co$_{85}$Fe$_{15}$ top layers, thus ruling out any
presence of spin-transfer torque acting on the spins in the
antiferromagnetic layer. Instead, our measurements show clear evidences for
the influence of Joule heating caused by the current. Moreover, results from
temperature-dependent measurements very much resemble those of the current
dependence, indicating that the effect of Joule heating plays a major role
in the current-in-plane spin-valve configurations. The results also suggest
that spin-transfer torques between ferromagnetic layers might still exist
and compete with the exchange bias at sufficiently high currents.
\end{abstract}


\maketitle

A current flowing through two nanoscale ferromagnets tends to align their
magnetic moments by a torque induced by the transfer of spin angular
momentum of the electrons polarized by one ferromagnet to the other
\cite{Slonczewski,Berger}. The technique based on this phenomenon for
controlling the state of ferromagnetic layers in spin-valve
\cite{Tsoi,Myers,Sun,Albert,Aoshima,Krivorotov} and magnetic tunneling
junction (MTJ) \cite{Huai,Fuchs,Yoshikawa,Assefa} structures has many
advantages expected to be useful in magnetic random access memory (MRAM)
technology. It has been reported recently \cite{Nunez,Wei,Urazhdin,Tang}
that a current could also induce a torque on the staggered moment of an
antiferromagnet and switch the direction of the exchange-bias field at the
antiferromagnet/ferromagnet interface. The switching of an antiferromagnet
(and therefore of the exchange bias field) even seems to appear at a lower
current density than for switching a ferromagnet \cite{Tang}.

N\'{u}\~{n}ez \textit{et al.} \cite{Nunez} proposed a theory based on a
model of 1D sandwich structure consisting of two antiferromagnets separated
by a paramagnetic spacer and predicted that currents could also alter the
micromagnetic state and induce a spin-transfer torque that acts on the
staggered moment of an antiferromagnet. Importantly, the authors found that
the critical current for switching an antiferromagnet is smaller than the
typical value for a ferromagnet because the spin transfer torques act
cooperatively throughout the entire antiferromagnet together with the
absence of shape anisotropy. Wei \textit{et al.} \cite{Wei} later reported
a variation of exchange bias by a high-density dc current injected from a
point contact into a spin valve; the exchange bias can increase or decrease
depending upon the current direction. The authors explained their results
in favor of the theory proposed by N\'{u}\~{n}ez \textit{et al.} \cite{Nunez},
i.e., electrons flowing from the ferromagnet into the antiferromagnet
induce torques on moments in the antiferromagetic matrix, altering its
magnetic configuration and favoring the parallel alignment of moments at the
ferromagnet/antiferromagnet interface, and therefore increase the
exchange-bias field, whereas electrons flowing in the opposite direction
tend to have the opposite effect. More convincing evidences for the effects
of spin polarized currents on an antiferromagnet in a spin-valve structure
have been also reported very recently by Urazhdin and Anthony
\cite{Urazhdin}.

In general, spin transfer or spin torque between the ferromagnetic layers in
a spin-valve (or MTJ) structure is expected when the current flows
perpendicular to layer planes (CPP) \cite{Slonczewski}. This also applies to
the case of antiferromagnetic spin torque because the current is needed to
pass through the antiferromagnetic layers \cite{Nunez,Wei}. Interestingly,
Tang \textit{et al.} \cite{Tang} have reported that dc currents flowing in
the layer planes (CIP) of an exchange-biased spin-valve can systematically
change the exchange bias. The authors also explained their results based on
an assumption that, together with the influence of the current-induced
field, electrons flowing into the antiferromagnetic layer from the
ferromagnetic one induce torques on the moments in the antiferromagnet. In
the present work, to avoid electrons flowing into the antiferromagnetic
layer in CIP measurements, we fabricated a spin-valve structure with an
extremely high resistance antiferromagnetic layer. However, we still observe
clear and systematic suppressions of the reversals of the pinned layer with
increasing current. Our results suggest that the effect caused by Joule
heating is significant and local spin torques between ferromagnetic layers
might still exist even in CIP configurations.

\begin{figure}[t!]
\includegraphics[width=6.cm]{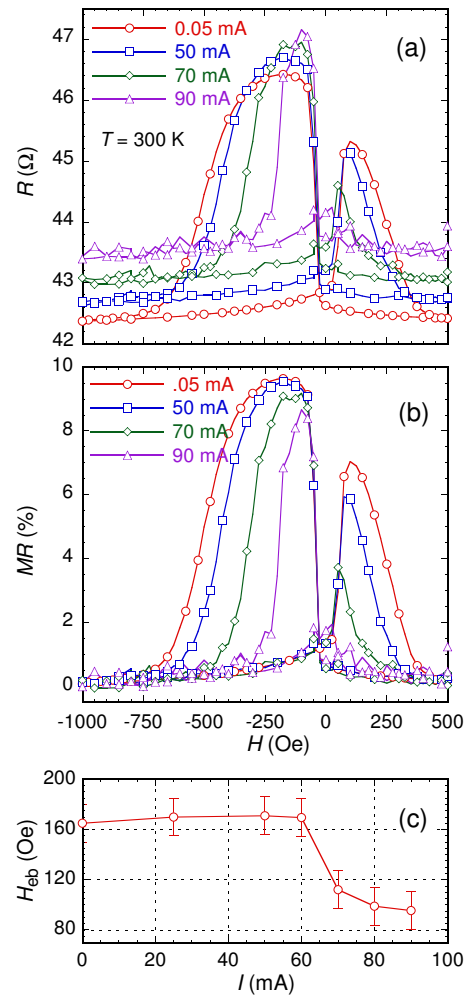}
\caption{(Color online) Variation of (a) the $R(H)$ and (b) corresponding
$MR(H)$ curves with current amplitude. Note that $R$, $MR$, and the
reversals of the pinned layer are all changed with current amplitude. In the
bottom panel (c), the $H_\mathrm{eb}$ data extracted from the $R(H)$ curves
are plotted against $I$.} \label{fig.1}
\end{figure}

The samples with a basic structure of
FeCo(2.4)/NiCoO(40)/FeCo(3.0)/Cu(3.8)/FeCo(4.5) (here, FeCo =
Fe$_{15}$Co$_{85}$, NiCoO = Ni$_{0.85}$Co$_{0.15}$O, and all the thicknesses
are in nm) were fabricated using a magnetron sputtering system (EDWARD AUTO
306) equipped with 3 targets, a dc and an rf source in the chamber with a
base vacuum of $8\times10^{-6}$ mbar. The FeCo and Cu layers were deposited
in Ar gas pressures of $5\times10^{-3}$ mbar and $3\times10^{-3}$ mbar,
respectively. The NiCoO antiferromagnetic layer was deposited from a
Ni$_{85}$Co$_{15}$ target in a $3\times10^{-3}$ mbar of a flowing mixture of
Ar and 20\% O$_{2}$. The resistance of our NiCoO films is so high, probably
more than 100 M$\Omega$, that we were unable to measure it accurately. An
external magnetic field of 500 Oe was aligned parallel to the substrate
plane during the deposition processes to create an initial exchange bias.
The quality of the films was examined by x-ray diffraction,
energy-dispersive x-ray spectroscopy, magnetization, and conductance
techniques. Resistance ($R$) and magnetoresistance
[$MR=(R_{\uparrow\downarrow}-R_{\upuparrows})\times100\%/R_{\upuparrows}$]
were measured on a $1\times5\ \mathrm{mm}^2$ sample using the standard
four-probe method by a Quantum Design PPMS 7100 with an ac current ($I$) at
a frequency $f=7.5$ Hz. The current was set to flow only during the duration
time $t_\mathrm{d}$ of measurement reading ($t_\mathrm{d}=0.3$ s was applied
for all measurements except those specified in Fig. 2). In all the
measurements, the external magnetic field ($H$), the sample's exchange-bias
and the injected current were always aligned in the same axis. Unless
otherwise specified in the temperature-dependent measurements (Fig. 4), all
the data were measured at $T=300$ K.

\begin{figure}[t!]
\includegraphics[width=6.cm]{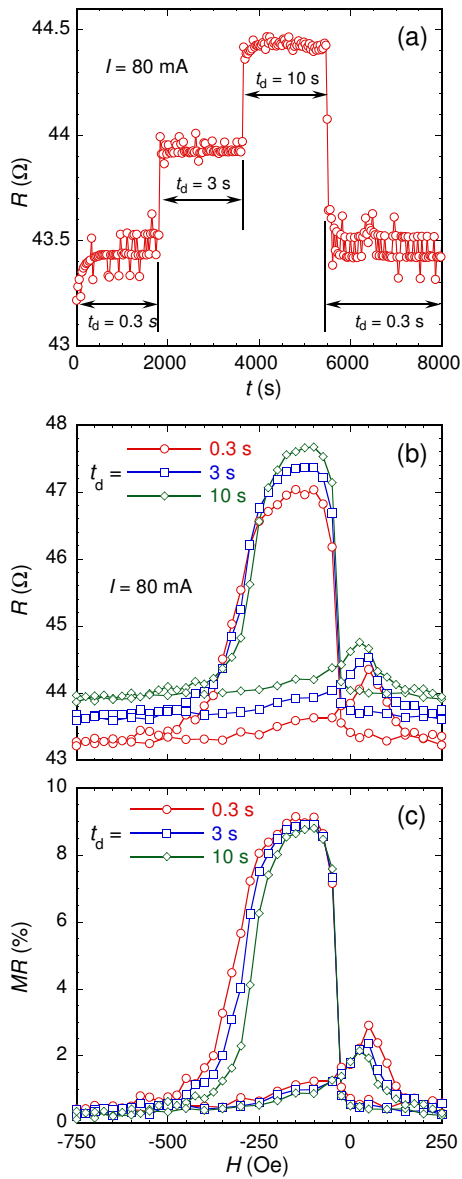}
\caption{(Color online) (a) $R$ is recorded vs. time while the duration time
for injecting a current $I=80$ mA is varied in a sequence $t_\mathrm{d}=0.3$
s (for 30 min) $\rightarrow$ 3 s (30 min) $\rightarrow$ 10 s (30 min)
$\rightarrow$ 0.3 s (53 min). The average interval between two current
injections is 30 s. (b) and (c) respectively display the $R(H)$ and
corresponding $MR(H)$ curves measured with $t_\mathrm{d}=0.3$, 3, and 10 s.
Note that $R$, $MR$, and the reversals of the pinned layer are all changed
with $t_\mathrm{d}$.} \label{fig.2}
\end{figure}

Representative $R(H)$ and $MR(H)$ curves are presented in Figs. 1(a) and
1(b), respectively. The reversal field of the free layer seems not to be
affected while that of the pinned layer (in both field sweeping directions)
is substantially suppressed. The $MR$ maximum on the $H>0$ branch is nearly
erased at $I=90$ mA when the reversal of the pinned layer occurs at a
magnetic field too close to that of the free layer. It is remarkable that
the results in Fig. 1 just resemble those previously observed in spin-valves
having a metallic antiferromagnetic layer where current flows to induce
torques on its magnetic moments therefore leading to a change
of the exchange bias \cite{Wei,Tang}. In our case, since the in-plane
current is confined to flow only in the FeCo(3.0)/Cu(3.8)/FeCo(4.5) top
layers, there would be no such current-induced spin torques in the
antiferromagnetic layer. Moreover, the suppressions of the reversals of the
pinned layer in our spin-valve may even not reflect a variation of exchange
bias. Since an exchange bias field causes a displacement of the $M(H)$ or
$R(H)$ curves to its reverse direction, a decrease of the exchange bias
should therefore always shift both of the reversals of the pinned layer
along the exchange bias direction. Our results in Fig. 1 show the
difference: with increasing current, the reversals of the pinned layer on
the $H>0$ and $H<0$ branches are oppositely shifted. In fact, by determining
the exchange bias field as
$H_\mathrm{eb}=-(H_\mathrm{r}^++H_\mathrm{r}^-)/2$ with $H_\mathrm{r}^+$ and
$H_\mathrm{r}^-$ are the fields at which the pinned layer reverses
completely on the $H>0$ and $H<0$ branches, respectively, we obtained
$H_\mathrm{eb}$ values of $\sim$170 Oe that is almost unchanged for currents
up to 60 mA, above which the exchange bias starts to decrease rapidly
[Fig. 1(c)]. This is a clear evidence that the changes of $H_\mathrm{r}^+$
and $H_\mathrm{r}^-$ are not related to a change of exchange bias, at least
for currents up to 60 mA, which is equivalent to a current density of
$\sim$$5\times10^5$ A/cm$^2$ flowing through the three top layers (note that
this current density is about more than 3 orders higher than the switching
current reported in Ref. \cite{Tang} for an antiferromagnet and still far
below the typical level for switching a ferromagnet). The magnetic field
generated by the ac current would cause an oscillation of the actual field
direction applied on the layers. If that oscillation is strong enough to
affect the reversals of the pinned layer, it should have already affected
the free layer that is more susceptible to magnetic field with a coercive
field $H_\mathrm{c}$ much smaller than both $H_\mathrm{r}^+$ and
$H_\mathrm{r}^-$. Clearly, our results do not support this reason.

\begin{figure}[t!]
\includegraphics[width=6.5cm]{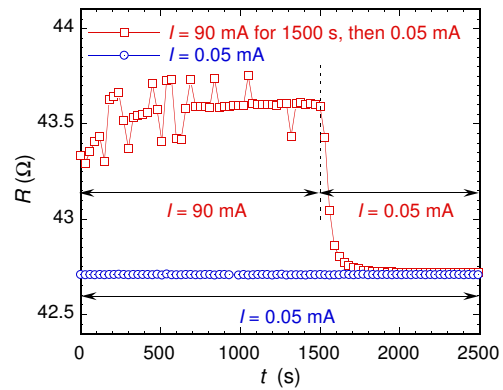}
\caption{(Color online) The influence of current amplitude on resistance.
The $\odot$-symbols: $R(t)$ at $I=0.05$ mA (negligible current heating). The
$\square$-symbols: $R(t)$ at $I=90$ mA and then at $I=0.05$ mA; the current
was switched at $t=1500$ s from 90 mA to 0.05 mA causing a change of
resistance from the heated state towards the unheated state.} \label{fig.3}
\end{figure}

The increases of resistance in both parallel and antiparallel states with
higher currents would suggest a possibility of Joule heating although the
current injection is turned on only for a short duration time $t_\mathrm{d}$
for resistance reading. The $R(H)$ curves in Fig. 1 were measured with
$t_\mathrm{d}=0.3$ s and the reading interval was kept unchanged for all the
measurements. Since the sample would take an amount of time much longer than
$t_\mathrm{d}$ to reach thermal equilibrium, an increase of $t_\mathrm{d}$
is thus expected to raise the sample's actual temperature that in turn
increases the resistance of the metallic layers. This possibility is
verified in Figs. 2(a) and 2(b) where an increase (decrease) of resistance
is always observed corresponding to an increase (decrease) of
$t_\mathrm{d}$. Figure 2(c) also shows that the reversals of the pinned
layer are monotonically suppressed with increasing $t_\mathrm{d}$, just as
increasing $I$. Obviously, the effects caused by current amplitudes and
duration times are qualitatively similar.

Another indication of the current heating effect is displayed in Fig. 3
where resistance data from two measurements are recorded with time. In one
measurement, $R(t)$ was measured at $I = 0.05$ mA and is considered as a
non-heating curve. On the other measurement, $I = 90$ mA was applied for the
first 1500 s and then the current was abruptly switched to 0.05 mA causing a
change of resistance towards the reference non-heating curve. This
observation corroborates the results in Fig. 1(a) that a higher current
gives a higher resistance value. Another noticeable feature recognized in
Fig. 3 is that when the current $I = 90$ mA was applied (at $t=0$ s) and
switched to 0.05 mA (at $t=1500$ s), correspondingly to a heating and a
cooling process respectively, the resistance takes a few minutes to reach
the equilibrium values.

\begin{figure}[t!]
\includegraphics[width=6.cm]{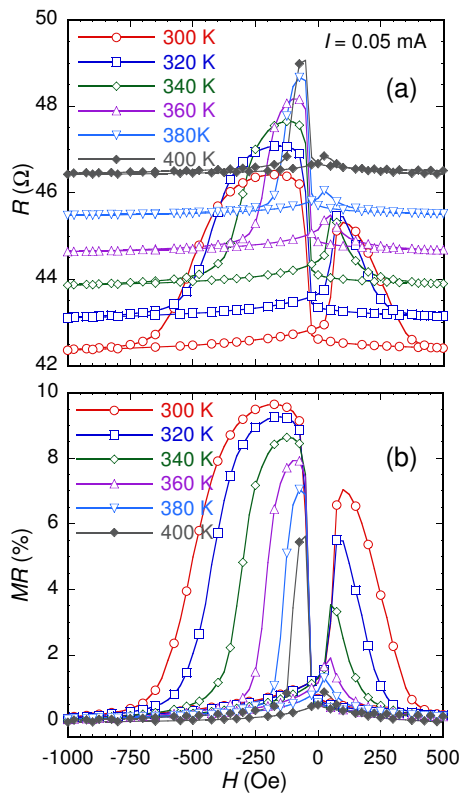}
\caption{(Color online) Variation of (a) the $R(H)$ and (b) corresponding
$MR(H)$ curves with temperature. Note that the changes in $R$, $MR$, and the
reversals of the pinned layer are qualitatively similar to those in Figs. 1
and 2.} \label{fig.4}
\end{figure}

As we have discussed, our results indicate that current heating plays an
important role in varying the magnetoresistance behavior of the CIP
spin-valve. In order to figure out whether the current can cause effects in
addition to the Joule heating, we carried out $R(H)$ measurements at
different temperatures from 300 K to 400 K. The influence of temperature, as
shown in Fig. 4, is qualitatively the same as that of current amplitude
(Fig. 1) or duration time (Fig. 2) except that the data are much less noisy.
A comparison of the results in Fig. 1 and Fig. 4 reveals an interesting
feature: heating is not the only direct effect caused by the current. The
$R(T)$ curves at $T=320$ K ($I=0.05$ mA) and $I=70$ mA ($T=300$ K) show
similar parallel and antiparallel resistances ($\sim$$43\ \Omega$ and
$\sim$$47\ \Omega$, respectively) but the exchange bias field is more
suppressed under the influence of current ($H_\mathrm{eb}\sim165$ Oe at
$I=0.05$ mA and $T=320$ K while $H_\mathrm{eb}\sim110$ Oe at $I=70$ mA and
$T=300$ K). On the other hand, the $R(T)$ curve at $I=70$ mA ($T=300$ K)
reveals a nearly similar $H_\mathrm{eb}\sim110$ Oe as the $T=340$ K
($I=0.05$ mA) curve, but the resistance values are smaller in both parallel
and antiparallel states ($\sim$43 $\Omega$ and $\sim$46.9 $\Omega$ compared
to $\sim$43.9 $\Omega$ and $\sim$47.7 $\Omega$, respectively). This implies
that the resistance (or exchange bias) is less raised (or more suppressed)
by currents than by purely temperature. We attribute this behavior to a spin
transfer between the two ferromagnetic layers. Even in a CIP configuration,
due to scattering processes, electrons are not confined to flow in one
specific layer but they still weave through and transfer spin angular
momentum between the two ferromagnetic layers. Such a spin transfer of
electrons in CIP configurations may not be as efficient as in CPP
configurations, but contributes to reducing spin scattering (thus lowering
resistance) and have an affection on the reversal of the pinned layer. It is
also possible that, at sufficiently high currents, the exchange bias is
directly affected by electron scattering at the interface between the pinned
ferromagnetic and the antiferromagnetic layer.

In conclusion, we have proved that Joule heating could play a major role in
the impact of in-plane current on the magnetotransport properties of a
spin-valve. The reversals of the pinned layer are suppressed even at low
currents when the exchange bias is not yet affected, probably by a decrease
of its coercive force with temperature. Not only do they generate heat,
in-plane currents also induce spin transfers between the ferromagnetic
layers that in turn compete with the exchange bias.

A part of this work was performed using facilities of the State Key Labs
(IMS, VAST, Vietnam). This work was also supported by MOST/KOSEF through
Quantum Photonic Science Research Center at Hanyang University (Seoul,
Korea). Two of us thank the University of Virginia for support.


\begin{references}

\bibitem{Slonczewski}
    J. C. Slonczewski, J. Magn. Magn. Mater. {\bf 159}, L1 (1996).
\bibitem{Berger}
    L. Berger, Phys. Rev. B {\bf 54}, 9353 (1996).
\bibitem{Tsoi}
    M. Tsoi, A. G. M. Jansen, J. Bass, W.-C. Chiang, M. Seck, V. Tsoi, and
    P. Wyder, Phys. Rev. Lett. {\bf 80}, 4281 (1998).
\bibitem{Myers}
    E. B. Myers, D. C. Ralph, J. A. Katine, R. N. Louie, and R. A. Buhrman,
    Science {\bf 285}, 867 (1999).
\bibitem{Sun}
    J. Z. Sun, J. Magn. Magn. Mater. {\bf 202}, 157 (1999).
\bibitem{Albert}
    F. J. Albert, J. A. Katine, R. A. Buhrman, and D. C. Ralph, Appl. Phys.
    Lett. {\bf 77}, 3809 (2000).
\bibitem{Aoshima}
    K.-I. Aoshima, N. Funabashi, K. Machida, Y. Miyamoto, N. Kawamura, K.
    Kuga, N. Shimidzu, and F. Sato, Appl. Phys. Lett. {\bf 91}, 052507
    (2007).
\bibitem{Krivorotov}
    I. N. Krivorotov, D. V. Berkov, N. L. Gorn, N. C. Emley, J. C. Sankey,
    D. C. Ralph, and R. A. Buhrman, Phys. Rev. B. {\bf 76}, 024418 (2007).
\bibitem{Huai}
    Y. Huai, F. Albert, P. Nguyen, M. Pakala, and T. Valet, Appl. Phys.
    Lett. {\bf 84}, 3118 (2004).
\bibitem{Fuchs}
    G. D. Fuchs, N. C. Emley, I. N. Krivorotov, P. M. Braganca, E. M. Ryan,
    S. I. Kiselev, J. C. Sankey, D. C. Ralph, R. A. Buhrman, and J. A.
    Katine, Appl. Phys. Lett. {\bf 85}, 1205 (2004).
\bibitem{Yoshikawa}
    M. Yoshikawa, T. Ueda, H. Aikawa, N. Shimomura, E. Kitagawa, M.
    Nakayama, T. Kai, K. Nishiyama, T. Nagase, T. Kishi, S. Ikegawa, and H.
    Yoda, J. App. Phys. {\bf 101}, 09A511 (2007).
\bibitem{Assefa}
    S. Assefa, J. Nowak, J. Z. Sun, E. O’Sullivan, S. Kanakasabapathy, W. J.
    Gallagher, Y. Nagamine, K. Tsunekawa, D. D. Djayaprawira, and N.
    Watanabe, J. App. Phys. {\bf 102}, 063901 (2007).
\bibitem{Krause}
    S. Krause, L. Berbil-Bautista, G. Herzog, M. Bode, R. Wiesendanger,
    Science {\bf 317}, 1537 (2007).
\bibitem{Nunez}
    A. S. N\'{u}\~{n}ez, R. A. Duine, P. Haney, and A. H. MacDonald, Phys.
    Rev. B {\bf 73}, 214426 (2006).
\bibitem{Wei}
    Z. Wei, A. Sharma, A. S. Nunez, P. M. Haney, R. A. Duine, J. Bass, A. H.
    MacDonald, and M. Tsoi, Phys. Rev. Lett. {\bf 98}, 116603 (2007).
\bibitem{Urazhdin}
    S. Urazhdin and N. Anthony, Phys. Rev. Lett. {\bf 99}, 046602 (2007).
\bibitem{Tang}
    X.-L. Tang, H.-W. Zhang, H. Su, Z.-Y. Zhong, and Y.-L. Jing, Appl. Phys.
    Lett. {\bf 91}, 122504 (2007).

\end{references}
\end{document}